\documentclass[%
 reprint,
superscriptaddress,
 amsmath,amssymb,
 aps,
]{revtex4-1}

\usepackage{graphicx}% Include figure files
\usepackage{dcolumn}% Align table columns on decimal point
\usepackage{bm}% bold math
\newcommand{\PP}{{\mathbb{P}}}

\newcommand{\tr}{{\text{Tr}}}

\newcommand{\hsgra}{\text{HSGRA}}

\begin{document}

\preprint{LMU-ASC 28/18}

\title{A Stringy theory in three dimensions and Massive Higher Spins}
\author{Evgeny Skvortsov}
\affiliation{%
Albert Einstein Institute,\\
Am M\"{u}hlenberg 1, D-14476, Potsdam-Golm, Germany 
}%
\affiliation{Lebedev Institute of Physics, \\
Leninsky ave. 53, 119991 Moscow, Russia}
\email{evgeny.skvortsov@aei.mpg.de}
\author{Tung Tran}
\affiliation{%
Albert Einstein Institute,\\
Am M\"{u}hlenberg 1, D-14476, Potsdam-Golm, Germany 
}%
\affiliation{%
 Arnold Sommerfeld Center for Theoretical Physics,\\
 Ludwig Maximilian University of Munich,\\
Theresienstr.  37, D-80333 M\"unchen, Germany
}%
\email{tung.tran@lmu.de}
\author{Mirian Tsulaia}
\affiliation{
Okinawa Institute of Science and Technology, \\ 1919-1 Tancha, Onna-son, Okinawa 904-0495, Japan}
\email{mirian.tsulaia@gmail.com}

\date{\today}% It is always \today, today,

\begin{abstract}
An example of a consistent theory with massive higher spin fields is constructed in flat space-time of dimension three. The action is written in the light-cone gauge. The theory has certain stringy features, e.g. its spectrum is unbounded in spin and mass, the theory admits Chan-Paton factors. The quartic and the higher tree-level amplitudes vanish, which softens the UV behaviour at the loop level and provides a new mechanism of how massive higher spin states can resolve the Quantum Gravity Problem. 

\end{abstract}

\maketitle

%\tableofcontents
%%%%%%%%%%%%%%%%%%%%%%%%%%%%%%%%%%%%%%%%%%%%%%%%%%%%%%%%%%
\section{\label{sec:intro}Introduction}
%%%%%%%%%%%%%%%%%%%%%%%%%%%%%%%%%%%%%%%%%%%%%%%%%%%%%%%%%%
Higher spin states are of crucial importance for constructing viable models of Quantum Gravity, at least within certain approaches. Indeed, lower spin extensions of gravity like super-gravities seem to suffer from UV divergences sooner or later \footnote{The jury is still out for the $\mathcal{N}=8$ SUGRA, see e.g. \cite{Bern:2018jmv} for the latest news.}. Therefore, higher spin states are needed, if there is a solution at all along these lines. The consistency of string theory relies on a very specific spectrum of higher spin states. Lastly, in $d\geq3$ even the smallest CFT's like the critical vector model have single-trace operators of arbitrarily high spin. Therefore, for any such CFT${}_d$ the holographically dual gravitation description in $AdS_{d+1}$  would have to contain higher spin states \footnote{We note that the strongly-coupled large-$N$ SYM theory has a large gap and there is a finite number of operators, including the stress-tensor, whose conformal dimensions are of order $1$. The rest of the spectrum, which contains higher spin states among others, corresponds to too heavy fields in $AdS_5$ and they effectively decouple in the zeroth approximation. However, they reappear at the next order. For the critical vector model \cite{Sezgin:2002rt, Klebanov:2002ja} the dual theory is supposed to be a \hsgra, see also \cite{Sundborg:2000wp}. However, the large-$N$ free and critical vector models do not have a large gap in the spectrum of single-trace operators and for this reason the dual theory does not satisfy the usual assumptions of locality and may not even exist \cite{Bekaert:2015tva,Maldacena:2015iua,Ponomarev:2017qab, Sleight:2017pcz}.}.

The general question we would like to address is whether or not there are consistent theories of quantum gravity that are much smaller than string theory and are as close to field theories as possible. These theories, if any, will have to contain infinitely many higher spin states and the spectrum has to be unbounded in spin \footnote{This is quite easy to show for massless higher spin fields \cite{Flato:1978qz,Berends:1984wp,Fradkin:1986ka,Maldacena:2011jn,Boulanger:2013zza}. For massive ones, the analysis of \cite{Metsaev:1991mt,Metsaev:1991nb,Ponomarev:2016lrm} can be straightforwardly generalized to the massive case with the same basic conclusion: once at least one higher spin field is present with a nontrivial self-interaction, the theory's spectrum has to be unbounded in spin.}. Having infinitely many fields in a theory is slightly outside the scope of the field theory in that the sum rules has to be prescribed by hand unless a more fundamental principle is understood, e.g. in string theory the worldsheet does the job. 

In looking for consistent theories with higher spin fields (ideally, for a Quantum Gravity) one has to start from a point in the theories' space that is far enough from string theory itself, the reason being that the Veniziano \cite{Veneziano:1968yb} and the like amplitudes seem to be quite unique \cite{Caron-Huot:2016icg}. We will use the light-front bootstrap --- the most general approach to local field theory where the constraints on the spectrum and amplitudes result from the closure of the Poincare (or any other) space-time symmetry algebra. The light-cone gauge allows one to directly deal with the dynamical degrees of freedom avoiding covariant descriptions that are usually ambiguous. For example, with the help of the light-cone approach one avoids dealing with the gauge symmetry, which is just a redundancy, whenever massless states are present.

One approach to the Quantum Gravity problem that systematically probes the most minimal higher spin extensions of gravity is Higher Spin Gravity (\hsgra). The main idea is to look for a completion of gravity with massless higher spin fields. The masslessness is expected to be equivalent to considering the high energy limit and the associated gauge symmetry is supposed to impose severe constraints on the structure of interactions and on possible counterterms. Since constructing a Quantum Gravity model has never been a simple task, many attempts to look for \hsgra's have faced numerous difficulties that can eventually be attributed to many no-go results against field theories with massless higher spin fields both in flat \footnote{See e.g. \cite{Weinberg:1964ew,Coleman:1967ad,Fotopoulos:2010ay,Bekaert:2010hp,Bekaert:2010hw} for a non-exhaustive list.} and anti-de Sitter spaces \footnote{See e.g. \cite{Dempster:2012vw,Maldacena:2011jn,Boulanger:2013zza,Bekaert:2015tva,Maldacena:2015iua,Ponomarev:2017qab, Sleight:2017pcz}, where some of the methods, ideas and results mimic those in flat space. For example, the Maldacena-Zhiboedov result is the $AdS/CFT$ analog of the Weinberg and Coleman-Mandula type theorems that completely fix (holographic) $S$-matrix whenever there are massless higher spin fields.}. 

The \hsgra{} programme has already been successful in giving a handful of classical theories that avoid the no-go results and provide rather simple models of quantum gravity, with the progress in proving the quantum consistency varying from model to model. At present, there are higher spin extensions of usual \cite{Blencowe:1988gj,Bergshoeff:1989ns,Campoleoni:2010zq,Henneaux:2010xg} and conformal gravity \cite{Pope:1989vj,Fradkin:1989xt,Grigoriev:2019xmp} in $3d$ that can all be formulated \cite{Grigoriev:2020lzu} as the Chern-Simons theory with a certain additional data \footnote{The additional data includes an embedding of the gravitational subalgebra into a given higher spin algebra, which determines the spectrum.}. There exists also a higher spin extension of the $4d$ conformal gravity \cite{Segal:2002gd,Tseytlin:2002gz,Bekaert:2010ky} with encouraging checks of the quantum corrections \cite{Beccaria:2016syk,Joung:2015eny}. Another useful model is the $4d$ chiral theory  \cite{Metsaev:1991mt,Metsaev:1991nb,Ponomarev:2016lrm} that exists both in flat and $AdS_4$ \cite{Metsaev:2018xip,Skvortsov:2018uru} backgrounds, being related to SDYM and QCD in the former \cite{Skvortsov:2020gpn} and to Chern-Simons Matter theories and dualities therein \cite{Giombi:2011kc, Maldacena:2012sf, Aharony:2012nh,Aharony:2015mjs,Karch:2016sxi,Seiberg:2016gmd} in the latter \cite{Skvortsov:2018uru}. The theory is one-loop finite \footnote{See \cite{Skvortsov:2018jea,Skvortsov:2020wtf,Skvortsov:2020gpn} and it also should be one-loop exact.}.

In the present paper we construct the first example of a theory with massive higher spin fields with the help of the light-front approach. We chose to do this in three dimensions which the lowest dimension where  massive higher spin fields have propagating degrees of freedom. At the same time, massless higher spin fields, including the graviton, do not have any local degrees of freedom in $3d$. Therefore, they do not exists in the light-cone approach. 

The outline is that we review the basics of the light-cone approach and then apply it to massive spinning fields in $3d$, where an exhaustive classification of cubic vertices has just been obtained by Metsaev \cite{Metsaev:2020gmb}.

%%%%%%%%%%%%%%%%%%%%%%%%%%%%%%%%%%%%%%%%%%%%%%%%%%%%%%%%%%
\section{\label{sec:hsgra}Light-Front Bootstrap}
%%%%%%%%%%%%%%%%%%%%%%%%%%%%%%%%%%%%%%%%%%%%%%%%%%%%%%%%%%
The main idea that dates back to Dirac \cite{Dirac:1949cp}, is that the combination of the relativity with the Hamiltonian dynamics implies that any classical or quantum field theory should deliver a realization of the space-time symmetry algebra, e.g. of the Poincare algebra if we are in flat space
\begin{align}
    [P_A,P_B]&=0\,,\\
    [J_{AB},P_C]&= P_A\eta_{BC}-P_B\eta_{AC}\,,\\
    [J_{AB},J_{CD}]&= J_{AD}\eta_{BC}+\text{3 more}\,.
\end{align}
Once a field theory is already known, the charges result from contracting the stress-tensor $T_{AB}$ with the Killing vector of the Poincare algebra and integrating over the Cauchy surface. Other way around, one can attempt to construct $P_A$ and $L_{AB}$ directly. Most of the generators, those that preserve the Cauchy surface, stay quadratic in the fields. What Dirac also noticed is that the number of the dynamical generators, i.e. those that are deformed by interactions, is minimal for the light-front quantization. They are the Hamiltonian and $(d-2)$ of the boost generators, $J^{a-}$, $A=a,+,-$. Therefore, one needs to solve   
\begin{align}
[J^{a-},P^{-}]&=0\,, & [J^{a-},J^{c-}]&=0\,.
\,.\label{stringeq}
\end{align}
These are exactly the equations that fix the critical dimension and intercept of string theory in the light-cone quantization \cite{Goddard:1973qh}. At the classical level, the second equation is a consequence of the first one. 

The light-cone gauge is a convenient choice to get to the bottom of a given theory, check its consistency and unitarity. What has not been much appreciated is that the light-front approach is an efficient tool to bootstrap new theories \cite{Bengtsson:1983pd,Bengtsson:1983pg,Metsaev:1991mt,Metsaev:1991nb,Metsaev:2005ar}. It also works in anti-de Sitter space \cite{Metsaev:2018xip} and for conformal field theories \cite{Skvortsov:2018uru}.

What one needs to do is to start out with a putative spectrum of fields and some basic interactions. Eq. \eqref{stringeq} will tell us if the spectrum needs to be enlarged and/or more interactions should be added. Eventually it fixes both the spectrum and all the couplings. To the lowest order the equations to be solved are  
\begin{align}\label{cubic}
    [H_2^{\phantom{a}}, J_3^{a-}]+[H_3^{\phantom{a}},J_2^{a-}]=0\,,
\end{align}
where $H\equiv P^-$ is the light-cone Hamiltonian and the subscript means the order of expansion. Free fields give $H_2$ and $J_2$ that are bilinear in the fields and are, of course, known \footnote{See e.g. \cite{Metsaev:2005ar}.}, e.g. 
\begin{align}
    H_2&= -\tfrac12\int d^{d-1}p\, \Phi^\dag_\mu(x^+,p)[{p_ap^a +m^2}]\Phi_\mu(x^+,p)\,.
\end{align}
In the light-cone gauge and in the momentum space each field is represented by $\Phi_{\mu}(x^+,p)$, where $\mu$ is an abstract spin label to distinguish between different irreducible fields, $x^+$ is the light-cone time and will be omitted in what follows \footnote{It is sufficient to check the relations of the Poincare algebra at $x^+=0$ and then evolve the generators according to the Hamiltonian $H$. }. Momentum $p$ consists of the transverse part $p^a$ and $p^+$, which we abbreviate as $\beta$. 

Eq. \eqref{cubic} fixes the cubic vertices, i.e. determines what are possible interactions among a given set of fields \footnote{See \cite{Bengtsson:1983pd,Bengtsson:1983pg} for the very first and \cite{Metsaev:2005ar} for the comprehensive list of cubic vertices.}. It does not yet fix the spectrum and the coupling constants in front of various independent cubic vertices. Usually, the decisive equation is the quartic one
\begin{align}\label{quartic}
    [H_3^{\phantom{a}}, J_3^{a-}]+[H_4^{\phantom{a}},J_2^{a-}]+[H_2^{\phantom{a}},J_4^{a-}]=0\,.
\end{align}
In most cases, it fixes the spectrum and the cubic couplings up to a few coupling constants. While in Yang-Mills-type theories the deformation stops at the quartic order, this is not so for generic gravitational theories. Therefore, finding $H_4$, $J^{a-}_4$ and the higher ones can be tedious. Nevertheless, one expects that \eqref{quartic} is constraining enough as to fix the spectrum of a theory and settle down the question of whether it exists or not. In some lucky cases, $H_4$, $J^{a-}_4$ and the higher ones vanish. Then, $H_3$ leads to an action that stops at the cubic level. One such example is the $4d$ chiral theory \cite{Metsaev:1991mt,Metsaev:1991nb,Ponomarev:2016lrm}. 

The light-front bootstrap should eventually be equivalent to the generalized unitarity methods, see e.g. \cite{Arkani-Hamed:2017jhn}, but it can be more handy and efficient sometimes.  

%%%%%%%%%%%%%%%%%%%%%%%%%%%%%%%%%%%%%%%%%%%%%%%%%%%%%%%%%%
\section{\label{sec:micro}Micro String Theory}
%%%%%%%%%%%%%%%%%%%%%%%%%%%%%%%%%%%%%%%%%%%%%%%%%%%%%%%%%%
In the paper we will look for the simplest solutions to the equations of the light-front bootstrap. Namely, we will look for theories with massive higher spin fields that do not require higher order corrections, i.e. we will attemt to solve
\begin{align}\label{chiral}
    [H_3^{\phantom{a}}, J_3^{a-}]=0\,.
\end{align}
In order to proceed we need the standard representation for $H_3$ and $J_3^{a-}$ in terms of densities $h_3$ and $j_3^{a-}$
\begin{align*}
    H_3&= \int d\Gamma_3\,  h_{3}(\PP,\beta_i,\mu_i)\, \tr \prod_{k=1}^{3} \Phi^\dag_{\mu_k}(p_k)\,,\\
    J_3^{a-}&= \int d\Gamma_3\,\left[j_{3}^{a-}-\frac13 h_{3}\sum_j\frac{\partial}{\partial p^a_i}\right]\tr\prod_{k=1}^{3} \Phi^\dag_{\mu_k}(p_k)\,,
\end{align*}
where $d\Gamma_n=\delta^{d-1}(\sum_j p_j)\prod_j d^{d-1}p_j$. The formulas are valid in any $d$. The sum over all $\mu_{1,2,3}$ that belong to the spectrum is implicit. The densities are functions of $\mu_i$, $\beta_i$, $\PP$, where $\PP^a=\tfrac13 \sum_j \check{\beta}_j p_j^a$ and $\check{\beta}_j=\beta_{j+1}-\beta_{j-1}$ modulo $3$. The trace $\tr$ is a trace over the color indices, if present. Now, with the help of the momentum conservation and of the canonical Dirac bracket
\begin{align}\label{Poisson}
    [\Phi_\mu(p),\Phi_{\mu'}(p')]&= \frac{\delta^{d-1}(p+p')}{2p^+} C_{\mu;\mu'}\,,
\end{align}
where $C_{\mu;\mu'}$ is a symmetric matrix, commutator \eqref{chiral} can be evaluated to 
\begin{align}
    [H_3^{\phantom{a}}, J_3^{a-}]=\int d\Gamma_4\, E^a(\mu_i,p_i,\beta_i)\,\tr \prod_{j=1}^{4} \Phi^\dag_{\mu_j}(p_j)\,.
\end{align}
The equation $E^a=0$ is the main equation to be solved and $E^a$ can be represented as ($\beta=\beta_1+\beta_2$)
\begin{widetext}
\begin{align*}
     E^a&= \sum h_3(\PP_{12};\beta_1,\beta_2,-\beta;\mu_1,\mu_2,\omega')\frac{C_{\omega;\omega'}}{2\beta}\left[j_3^{a-}(\PP_{34};\beta_3,\beta_4,\beta;\mu_3,\mu_4,\omega)-\frac13(\beta_3-\beta_4)\frac{\partial}{\partial \PP^a}h_3(\PP_{34};\beta_3,\beta_4,\beta;\mu_3,\mu_4,\omega)\right]
\end{align*}
\end{widetext}
The $\sum$ corresponds to summation over (i) the exchanged states $\omega$, $\omega'$; (ii) permutations of the external legs, e.g. over the cyclic permutations of the arguments if the trace $\tr$ over the color indices is retained or over all permutations if there are no color indices and the fields on the external lines are the same. 

Our study \footnote{A thorough discussion of all possible cases will be given elsewhere. While preparing this manuscript we learned about the 2nd version of \cite{Metsaev:2020gmb}, whose Appendix B has a significant overlap with our results.} shows that generic interactions cannot satisfy \eqref{chiral} or $E^a=0$ for a very simple but technical reason. Therefore, we turn to the interactions that are specific to three dimensions. In $3d$ the label $\mu$ is a pair $(s,m)$ where $s$ is the spin, $s\geq0$, and $m$ is a mass, i.e. we have $\phi_{s,m}(p)$. For $s>0$ the two signs of the mass, $m>0$ and $m<0$, correspond to different irreducible fields. It is more convenient to introduce complex conjugate fields
\begin{align}
    \Phi_{\lambda, m}^\dag(p)=\Phi_{-\lambda,m}(-p)
\end{align}
that obey \eqref{Poisson} with $C_{\lambda, m; \lambda', m'}=\delta_{m,m'}\delta_{\lambda+\lambda',0}$. The scalar field $\Phi_{0,m}$ is real. A complete classification of all cubic interaction vertices in $3d$ is available \footnote{This was nicely done recently by Metsaev in \cite{Metsaev:2020gmb}. Note that the results of the present paper are inspired by the dimensional reduction. }. 

One of the main results of the present paper is that the following Hamiltonian solves $E^a=0$ and, hence, gives an example of a consistent theory:
\begin{align*}
    h_3= \sum_{\lambda_i=-\infty}^{+\infty}\sum_{k_i}  C(k_i,\lambda_i) V(\PP,\beta_i,k_i,\lambda_i) \,,
\end{align*}
where $i=1,2,3$. The coupling constants are 
\begin{align}
    C&=\frac{\delta_{\sum_i k_i\epsilon_i,0} }{\Gamma[\Lambda]}\,, \qquad \Lambda=\sum_i\lambda_i\,,
\end{align}
and the vertex reads \footnote{These vertices are related by a field-redefinition to those in the classification of \cite{Metsaev:2020gmb}, namely to the case $Ib$ there and follow from the dimensional reduction as well. }
\begin{align}
    V&=(\mathbb{P}+\mathbb{P}_{\lambda})^{\Lambda} \prod_i \beta_i ^{-\lambda_i} \,.
\end{align}
Here we also define 
\begin{align}
   \mathbb{P}_{\lambda}&=\frac{i}3 m \sum \check{\beta}_j \epsilon_j k_j\,, \qquad \epsilon_i=\mathrm{sign}(\lambda_i)\,.
\end{align}
The spectrum of the theory contains all spins $s=0,1,2,3,...$ and all masses that are integer multiples, $k\,m$, of some basic mass $m$. It is also possible to introduce color factors, $\Phi\equiv \Phi_\alpha T^\alpha$, by making $\Phi$ matrix-valued. It is easy to see that $U(N)$, $O(N)$ and $USp(N)$ gaugings are possible \footnote{This is identical to \cite{Metsaev:1991nb,Skvortsov:2020wtf}, see also \cite{Konstein:1989ij} for the first occurrence of Chan-Paton factors in the higher spin context.}.  

What should be remembered is that, even though free massive fields can be obtained via dimensional reduction from the massless ones, this is not so for interactions. There are many more vertices among massive higher spin fields than can be obtained via a compactification. Nevertheless, the present theories are relatives of the $4d$ chiral theory. Indeed, one can obtain the Hamiltonian above as a compactification on the circle along $x^3$ via \footnote{This same idea has just been explained in the 2nd version of \cite{Metsaev:2020gmb}.}
\begin{align*}
    \Psi_\lambda(p,x^3)&=\sum_{k} \exp[ikmx^3\mathrm{sign}(\lambda)]\Phi_{\lambda,m k}(p)\,.
\end{align*}
Due to the somewhat formal nature of such a compactification, e.g. there are no massless fields with $s>0$ in $3d$ \footnote{Note, however, that in the light-cone gauge the $m\rightarrow0$ limit of the $3d$ massive spin-$s$ field gives the $3d$ massless scalar field. }, it is necessary to check the closure of the algebra again. In fact, it is more illuminating to start out in $3d$ and investigate the constraints imposed by $E^a=0$. 

Generic interactions, i.e. without fine-tuned masses, require higher orders and cannot solve $E^a=0$. Non-generic cubic interactions require \footnote{The conditions of this kind first appeared in \cite{Fotopoulos:2009iw} and play a crucial role in $3d$ \cite{Metsaev:2020gmb}.}
\begin{align}
    \sum_j \epsilon_j m_j&=0\,, && \epsilon_j\in \{\pm 1\}\,.
\end{align}
Therefore, masses tend to form a lattice. Further constraints follow from $E^a=0$ that requires every spin-$s$ 'exchange' that contributes to $E^a$ to be a member of a family of exchanges that start at the lowest possible spin and ends at the highest possible spin. This implies a formation of Regge-like trajectories. Therefore, certain crucial features of string theory are already visible in the smallest theories with higher spin states.

A straightforward computation shows that the quartic amplitude vanishes on-shell and the sum over the KK-modes does not create any problem due to the conservation of the lattice momentum \footnote{One needs to use the identities collected in \cite{Skvortsov:2020wtf} and adapt them to some of the momenta components being discrete.}. Using the Berends-Giele currents \cite{Berends:1987me} as in \cite{Skvortsov:2018jea,Skvortsov:2020wtf} one can show that the higher tree-level amplitudes vanish as well, i.e. we have $\mathcal{A}_{n,\text{tree}}(p_1,...,p_n)=0$ \footnote{This computation parrots the one in \cite{Skvortsov:2020wtf} and is not worth discussing in detail.}. 

The vanishing of the tree-level
amplitudes $\mathcal{A}_{n,\text{tree}}$ has to soften the UV-behaviour of the loop corrections \footnote{The computations are lengthy and will be presented elsewhere}. Therefore, as in the $4d$ chiral theory, we expect the $n$-point one-loop amplitudes to be UV-finite \cite{Skvortsov:2018jea,Skvortsov:2020wtf,Skvortsov:2020gpn}. We note in passing that in the chiral theory the one-loop amplitudes are closely related to those of QCD and SDYM \cite{Skvortsov:2020wtf,Skvortsov:2020gpn}. The UV-finiteness strengthens the importance of higher spin states present in the form of Regge-like trajectories for the UV-consistency of the theory. 

In addition, each loop diagram is accompanied by a purely numerical and divergent factor $\nu=\sum_\lambda 1$. In the $4d$ case the Weinberg low energy theorem instructs us to set $\nu=0$, which is also consistent with a web of results on one-loop determinants \cite{Gopakumar:2011qs,Tseytlin:2013jya,Giombi:2013fka,Giombi:2014yra,Beccaria:2014jxa,Beccaria:2014xda,Beccaria:2015vaa,Gunaydin:2016amv,Bae:2016rgm,Skvortsov:2017ldz} and especially with  \cite{Beccaria:2015vaa}. For massive higher spin fields we are not obliged to set $\nu=0$ and it would be interesting to see if there are other consistent choices. 

%%%%%%%%%%%%%%%%%%%%%%%%%%%%%%%%%%%%%%%%%%%%%%%%%%%%%%%%%%
\section{\label{sec:conc}Conclusions}
%%%%%%%%%%%%%%%%%%%%%%%%%%%%%%%%%%%%%%%%%%%%%%%%%%%%%%%%%%
We have looked for the most minimal theories (those that do not need higher corrections, $H_n$, $n\geq 4$, to the Hamiltonian) with massive higher spin fields in $3d$, but even in this case the complete classification of solutions to the light-front equations is not yet known \footnote{See \cite{Ponomarev:2016lrm,Ponomarev:2017nrr} for the systematic study of possible solutions in the $4d$ massless case }. In addition, there should exist theories that do not stop at the cubic order and where the present one is as a closed subsector \footnote{We note that in the massless case, the $4d$ chiral theory is unique and cannot be extended to a perturbatively local higher spin theory as the latter does not exist. This is true both in flat \cite{Bekaert:2010hp,Fotopoulos:2010ay,Ponomarev:2017nrr,Taronna:2017wbx,Roiban:2017iqg} and anti-de Sitter cases \cite{Dempster:2012vw,Bekaert:2015tva,Maldacena:2015iua,Ponomarev:2017qab, Sleight:2017pcz}.}.  Massive higher spin fields can resolve some of the singularities faced for massless fields \footnote{For example, the no-go observed in \cite{Ponomarev:2017nrr} for the massless case is not applicable to the present case since there are many new interactions among massive $3d$ spinning fields than can be obtained by the dimensional reduction}. In particular, massive higher spin fields in Minkowski space can, in some sense, model massless higher spin fields in $AdS$ since the latter also have mass-like terms. 

There is one more consistent higher-spin theory that immediately follows from \cite{Ponomarev:2016lrm}: its spectrum consists of spin-two and spin-$s$ fields (here $s$ is fixed) with KK-masses $k\, m$. It results from reducing a consistent gravitational coupling of the $4d$ massless spin-$s$ field that includes the spin-two self-coupling as well. As is known \cite{Fradkin:1982kf}, the consistency of dimensionally reduced theories is a nontrivial issue and requires careful regularization.

Another interesting application is motivated by the zoo of the massive spin-two theories in $3d$, see e.g. \cite{Afshar:2019npk} for a review of all known cases. It is obvious that many of the ideas and approaches admit a generalization to higher spin fields. It would be important to explore this direction further and to construct Lorentz covariant examples of theories with massive higher spin fields. In particular, we cannot see in the light-cone gauge if the theory discussed above can be coupled to $3d$ gravity since the latter has no local degrees of freedom.  

It would also be important to pursue the programme of bootstrapping theories with massive higher spin fields further. In particular, the most interesting applications are expected to be in four-dimensions. Here, one should start out with a graviton, as a massless spin-two field, and at least one massive higher spin field, assuming them to be minimally coupled. Our preliminary considerations indicate that there should exist such a theory with a graviton and massive higher spin fields \footnote{In particular, it should be possible to avoid some of the problems observed previously for massless higher spin fields \cite{Ponomarev:2017nrr,Taronna:2017wbx,Roiban:2017iqg}. } 

More generally, one should investigate further if there are consistent theories with higher spin fields that are much smaller than string theory, which should shed more light on the Quantum Gravity Problem. One advantage of the bottom-up approach, in particular of the light-front one, is that it should eventually be possible to chart out the landscape of all consistent theories.   

\begin{acknowledgments}
We are grateful to Sadik Deger, Axel Kleinschmidt, Ruslan Metsaev, Yasha Neiman, Radu Roiban and to Arkady Tseytlin for the very useful discussions. The work of E.S. was supported by the Russian Science Foundation grant 18-72-10123 in association with the Lebedev Physical Institute. The work of T.T. is supported by the International Max Planck Research School for Mathematical and Physical Aspects of Gravitation, Cosmology and Quantum Field Theory. The work of M.T. was supported by the Quantum Gravity Unit of the Okinawa Institute of Science and Technology Graduate University (OIST).
\end{acknowledgments}
\appendix

\end{document}